\documentclass{article}

\input{tcilatex}

\begin{document}

\textbf{Faur\'{e} and Buzdin Reply: }In our work \cite{Faure and Buzdin}, we
theoretically studied the evolution of the domain structure in a film of
superconducting ferromagnet (SFM) at the transition from the normal to the
superconducting state. We obtained rather rich variety of scenarios
depending on the parameters of the system. For $\lambda >\widetilde{w}%
/\left( 8\pi \right) $ (where $\lambda $ is the London penetration depth and 
$\widetilde{w}$ is the effective domain wall with), we demonstrated that
superconductivity decreases the domain size. The problem of the domain
structure in SFM was previously addressed by Sonin \cite{Sonin1} who studied
"...material, in which the magnetic transition occurs earlier, i.e. at a
higher temperature, than the onset of superconductivity." He specified that
"... our goal is the domain structure due to magnetostatic fields generated
by nonzero average bulk magnetization M. In this case the domain size
depends on the sample size" and concluded that the domain structure is
absent in the Meissner state. Our analysis precisely referred to this
situation and the free energy in SFM is the same as (1) from \cite{Sonin1}.
We showed that such a scenario is realized only for 
\begin{equation}
\lambda <\widetilde{w}/\left( 8\pi \right) .  \label{1}
\end{equation}%
We think this is an important point that prevents the misinterpretation of
the results of \cite{Sonin1} and that was underlined in our article. In
fact, this condition simply means that the ferromagnetic induction in SFM is
completely screened and the total magnetic induction is zero. Therefore,
there is no stray field in that case and obviously no reason for the domain
structure formation in SFM (not only in the plate-like geometry but also in
any ellipsoidal shape).

In his Comment \cite{Sonin2} Sonin claims "...Faur\'{e} and Buzdin [1]
considered the intrinsic domain structure, which was analyzed by Krey [6]
more than 30 years ago." This is an error. Our main result is the general
expression given by formula (3) in \cite{Faure and Buzdin} for the energy of
the domain structure in a SFM film of arbitrary thickness $2L_{z}$ (and not
an intrinsic domain structure). The minimization of (3) from \cite{Faure and
Buzdin} gives the equilibrium domain width $\ell $. Performing the
corresponding elementary calculations in the limit $\lambda <<\left( 
\widetilde{w}L_{z}^{3}\right) ^{1/4}$, it reads 
\begin{equation}
\ell =\left( \frac{3\lambda ^{2}\widetilde{w}}{2\pi }\right) ^{1/3}\left[ 1-%
\frac{7\varsigma (3)}{\pi ^{8/3}}\left( \frac{2}{3}\right) ^{1/3}\frac{%
\lambda ^{4/3}}{L_{z}\widetilde{w}^{1/3}}\right] .  \label{2}
\end{equation}%
The expression (7) from \cite{Faure and Buzdin} is simply the leading term
into (\ref{2}), when the volume contribution becomes dominant, and not at
all the main result of our work as Sonin presents. It is not astonishing
that this case corresponds to the result of Krey for the volume domain
structure and we are grateful to Sonin for this reference. Evidently, we
must retrieve the bulk result (which is the intrinsic domain structure) for
large film thicknesses $L_{z}$ . Note that we have retained the second
corrective term in (\ref{2}), which naturally depends on the film thickness
and unambiguously shows the general character of our approach . It is also
absolutely clear from formulas (3), (5-6) and Fig. 2 from \cite{Faure and
Buzdin} that we considered the domain structure in a thin film and not the
intrinsic domain structure. For example, the formula (6) from \cite{Faure
and Buzdin} for the domain width $\ell $ has nothing to do with the
intrinsic domain structure. Therefore, the problem we studied is exactly the
same as the one studied by Sonin \cite{Sonin1} and we just indicated in \cite%
{Faure and Buzdin} at what condition we obtain qualitatively different
results from that of \cite{Sonin1}. Besides, this condition should have been
specified in \cite{Sonin1} and it is regrettable that it was not.

It follows from the last part of Sonin's Comment \cite{Sonin2} that he in
fact agrees with the condition of the absence of domains $\lambda <%
\widetilde{w}/\left( 8\pi \right) .$ If the magnetic anisotropy is related
to the magnetodipole interaction, then the quality factor is of the order of
unity and effective domain width is close to the real one. Moreover, the
divergency of the London penetration depth at $T\rightarrow $ $T_{c}$
implies that the situation $\lambda (T)>\widetilde{w}/\left( 8\pi \right) $
is always realized near $T_{c}$. Besides, the domain structure of SFM URhGe
in the normal state was very recently studied with the help of SQUID
microscope \cite{Klaus Hasselbach}. In a film of $0.4$ $mm$, domains were
observed whose width was of the order of $20\mu m$. This gives us the direct
estimate of the effective domain width $\widetilde{w}\sim 1$ $\mu m$. On the
the other hand, in this SFM \cite{Aoki}, the London penetration depth $%
\lambda \sim 9000$ $\mathring{A}$. Consequently, the condition $\lambda >>$ $%
\widetilde{w}/\left( 8\pi \right) \ $is fullfilled. Therefore, we expect a
contraction of the domain structure below the critical temperature in URhGe.

M. Faur\'{e}$^{1}$ and A. Buzdin$^{1,2}$.

\bigskip$^{2}$ Institut Universitaire de France, Paris, France


\begin{thebibliography}{9}
\bibitem{Faure and Buzdin} M. Faur\'{e} and A. I. Buzdin, Phys. Rev. Lett. 
\textbf{94}, 187202 (2005).

\bibitem{Sonin1} E. B. Sonin, Phys. Rev. B \textbf{66}, 136501 (2002).

\bibitem{Sonin2} E. B. Sonin, preceding Comment.

\bibitem{Krey} U. Krey, Intern. J. Magnetism, \textbf{3}, 65 (1972).

\bibitem{Klaus Hasselbach} K. Hasselbach and V. Dolocan, private
communication.

\bibitem{Aoki} D. Aoki et al., Nature (London) \textbf{413,} 613 (2001).
\end{thebibliography}
\end{document}